\documentclass[aps,prb,twocolumn,showpacs]{revtex4-1}
\usepackage{amsmath}

\usepackage{graphicx}

\begin{document}

\title{The Berry phase and the phase of the Shubnikov-de Haas oscillations
in three-dimensional topological insulators}

\author{G.~P.~Mikitik}

\author{Yu.~V.~Sharlai}

\affiliation{B.~Verkin Institute for Low Temperature Physics \&
  Engineering, Ukrainian Academy of Sciences,
   Kharkov 61103, Ukraine}

\date{\today}

\begin{abstract} Within the semiclassical approach, we
calculate contributions of the Berry phase and of the Zeeman
coupling of an electron moment with the magnetic field to the
phase of the Shubnikov - de Haas oscillations for the surface
electrons in the Bi$_2$X$_3$ family of three-dimensional
topological insulators (X stands for Te or Se). We also discuss a
relation of the obtained results with published experimental data
on the Shubnikov-de Haas oscillations for this family of
topological insulators.
\end{abstract}


\maketitle

Three-dimensional topological insulators (TI) attract considerable
interest as a new state of solids. \cite{rmp10} These materials
are insulating in the bulk and are metallic on their surfaces.
Recently, a new important class of these materials was discovered
that consists of Bi$_2$X$_3$ compounds where X stands for Te or
Se. \cite{nat1,zhang} The angle-resolved photoemission
spectroscopy revealed that the energy bands of the surface
electrons are practically linear in the wave vector ${\bf k}$ in
these compounds, forming a massless Dirac cone in their
surfaces.\cite{nat1,sci,hsieh} It is necessary to emphasize that
although a similar Dirac cone exists in graphene, the surface
electron states in the TI essentially differ from the electron
states in graphene. The surface states of the TI have no inversion
symmetry, and the spin-orbit coupling is not weak in these
crystals. These features of the TI lead to a locking of electron
spins with ${\bf k}$ for the surface states, giving rise to Dirac
fermions without spin degeneracy.

Recently, the quantum oscillations of conductivity in the magnetic
field (the Shubnikov-de Haas oscillations) were observed in
(Bi$_{1-x}$Sb$_x$)$_2$Se$_3$, \cite{nat} Bi$_2$Te$_3$,
\cite{ong,xiu} Bi$_2$Te$_2$Se, \cite{taskin,xiong}
Bi$_{1.5}$Sb$_{0.5}$Te$_{1.7}$Se$_{1.3}$, \cite{taskin11} and
Bi$_2$Se$_3$. \cite{sacepe} It was found that the observed
oscillations result from the surface electrons of these TI. For
such a two-dimensional metal the part of the conductivity
describing these oscillations in the semiclassical limit,
$\delta\sigma_{xx}$, has the form: \cite{LP-k,Sh,gus,shen}
 \begin{equation}\label{SH}
\delta\sigma_{xx}(1/H)=\sum_{l=1}^{\infty}A_l
\cos\left[l\left(2\pi\frac{F}{H}+ \varphi \right)\right],
 \end{equation}
where $A_l$ are positive amplitudes of the harmonics of the
periodic in $1/H$ signal, $F$ and $\varphi$ are the frequency and
the phase of the oscillations, the $x$-$y$ plane coincides with
the surface of TI, and the magnetic field $H=H_z$ is perpendicular
to this surface. The frequency $F$ is determined by the area
$S(\varepsilon_F$) of the closed orbit of an electron with the
Fermi energy $\varepsilon_F$ in the space of the wave vectors
${\bf k}$, $2\pi F=\hbar c S(\varepsilon_F)/e$, \cite{LP-k,Sh}
where $e$ is the absolute value of the electron charge. The phase
$\varphi$ is given by the constant $\gamma$, $\varphi= -2\pi
\gamma$, \cite{shen} that enters the semiclassical quantization
condition for the electron energy in the magnetic field,
Eq.~(\ref{1}). Both $F$ and $\varphi$ can be experimentally
determined by Fourier analysis of $\delta \sigma_{xx} (1/H)$.
\cite{K} It was found in the experiments \cite{xiu,xiong,sacepe}
that the phase of the oscillation, $\varphi$, practically
coincides with the appropriate phase for the electrons in
graphene, i.e., $\varphi\approx 0$. On the other hands, this phase
measured in other experimental investigations
\cite{nat,ong,taskin,taskin11} generally differs both from the
phase for the usual two-dimensional electron gas ($\varphi=-\pi$)
and from the phase in graphene. As will become clear below, the
constant $\gamma$ (i.e., $\varphi$) depends on the so-called Berry
phase \cite{berry} of electrons and on the Zeeman coupling of an
electron moment with the magnetic field. In
Refs.~\onlinecite{nat,taskin,ong} it is the Zeeman coupling that
was discussed as a possible source of the nonzero $\varphi$.
Taskin and Ando \cite{taskin1} suggested that a deviation of the
dispersion relation $\varepsilon({\bf k})$ for the surface charge
carriers from an ideal cone can shift their Berry phase $\Phi_B$
from its usual value $\pi$ characteristic of the Dirac electrons
and hence can change the phase of the oscillations $\varphi$, too.
In this paper, using the semiclassical approach, we calculate both
the Berry phase and a contribution of the Zeeman coupling to the
constant $\gamma$ for the surface electrons in the Bi$_2$X$_3$
family of TI. We also discuss a relation of these results with the
published experimental data.

The semiclassical quantization condition for the energy levels
$\varepsilon_n$ of an electron in the magnetic field $H$ reads:
\cite{LP,Sh}
 \begin{equation}\label{1}
 S(\varepsilon_n)=\frac{2\pi e H}{\hbar c}(n+\gamma ),
 \end{equation}
where  ${n}$ is a large integer. The quantization condition
(\ref{1}) is obtainable from the one-band Hamiltonian
\cite{blount} of a semiclassical electron in the magnetic field.
This Hamiltonian $\hat H$ can be schematically represented in the
form:
\begin{equation}\label{2}
 \hat H= \varepsilon_0({\bf k})-\frac{eH}{c}
 (\mu_0^{(1)}+\mu_0^{(2)}),
 \end{equation}
where the subscript $0$ marks a two-dimensional electron-surface
band under study, the term
\begin{eqnarray}\label{4}
 \mu_{0}^{(1)}({\bf k})&=&[{\bf v}_0\times {\bf \Omega}_0]_z
\end{eqnarray}
is proportional to the intraband part of the orbital electron
moment, ${\bf v}_0= (1/\hbar)(\partial \varepsilon_0/\partial {\bf
k})$ is the electron velocity,
\begin{equation} \label{6}
{\bf \Omega}_0({\bf k})= i\int u^{*}_{{\bf k},0}({\bf r}) \frac
\partial {\partial {\bf k}}  u_{{\bf k},0}({\bf r})d{\bf r},
\end{equation}
and $\mu_0^{(2)}$ is the sum of the $z$ components of the
interband orbital moment of the electron and of its spin. In
Eq.~(\ref{6}) the integration is carried out over a unit cell of
the crystal lattice, and $u_{{\bf k},0}({\bf r})$ is the periodic
factor in the Bloch wave function of the $0$-th band,
\[
\psi_{{\bf k},0}= {\rm exp}(i {\bf kr})u_{{\bf k},0}.
\]

Before proceeding to TI, let us briefly review possible situations
that can occur in crystals and that have an effect on the form of
Eq.~(\ref{1}). If the spin-orbit interaction in a crystal is
negligible, and if the electron states under study have the
inversion symmetry, the constant $\gamma$ is given by the
formula:\cite{prl}
 \begin{equation}\label{2a}
 \gamma = \frac{1}{2}-\frac{\Phi_B}{2\pi}\ ,
 \end{equation}
where $\Phi_B$ is the Berry phase of the electron orbit $\Gamma$,
\begin{equation}\label{3a}
 \Phi_B= \oint_{\Gamma} {\bf \Omega}_0 d {\bf k},
 \end{equation}
$d{\bf k}$ $\equiv d \kappa [ {\bf i}_z \times {\bf v_0} ] /|{\bf
v_0}|$; ${d \kappa }$ is the length of an infinitesimal element of
the orbit ${\Gamma }$, and ${{\bf i}_z}$ is the unit vector
parallel to ${\bf H}$. It is this situation that occurs in
graphene. The part $\Phi_B/2\pi$ of Eq.~(\ref{2a}) results from
the term $\mu_0^{(1)}$ in Hamiltonian (\ref{2}). The term
$\mu_0^{(2)}$ is identically equal to zero in this case (if one
neglects the electron spin). As was shown in our paper, \cite{prl}
the Berry phase $\Phi_B$ of an electron is always equal to $\pm
\pi$ when it moves around a Dirac point, i.e., if the electron
orbit $\Gamma$ surrounds this point. Otherwise, one has
$\Phi_B=0$. Importantly, the Berry phase does not depend on the
shape and the size of the electron orbit since the inversion and
time-reversal symmetries lead to ${\rm rot}{\bf \Omega}_0=0$
everywhere except the Dirac point. In other words, the result
$\Phi_B=\pm \pi$ remains true at {\it any dependence}
$\varepsilon_0({\bf k})$ {\it in the vicinity of the orbit}, and
$\gamma$ can take on only the universal values $0$ or $1/2$.

If the spin-orbit interaction is not weak, but still there is an
inversion symmetry for the electron states, all these states are
double degenerate in spin. In this case one has $\gamma=1/2$, but
the quantization rule (\ref{1}) contains an additional term
associated with the so-called electron $g$ factor, $g$, i.e., in
this case $\gamma \to 1/2 \pm gm_*/4m$ where $m$ is the electron
mass and $m_*$ is its cyclotron mass. One part of this $g$ factor
is still determined by the Berry phase, whereas its second part
depends on the term $\mu_0^{(2)}$ which is not vanish now. The
complete theory of the $g$ factor for electrons in metals was
presented in our papers. \cite{g-fac,g-fac1} It is important that
if the strength of the spin-orbit interaction is small, the $g$
factor approaches a limiting form that leads to the same electron
spectrum described by Eq.~(\ref{1}) with $\gamma=1/2$ or $0$ as in
the absence of this interaction. \cite{zhetf}

When the inversion symmetry is absent, formulas for the constant
$\gamma$ can be derived by a simple modification of the
expressions given in Refs.~\onlinecite{g-fac,g-fac1}. In
particular, in the case of TI when the spin-orbit interaction is
not weak, and hence when the surface electron states are not
degenerate in the electron spin, we obtain
\begin{equation} \label{7}
\gamma -\frac{1}{2}=-\frac{1}{2 \pi} \oint_{\Gamma} {\bf \Omega}_0
d {\bf k} -\frac{1}{2 \pi } \oint_{\Gamma } \frac{\mu_0^{(2)}({\bf
k})} {v_0({\bf k})} d \kappa.
\end{equation}
Here the first integral coincides with $-\Phi_B/2\pi$, while the
second integral can be considered as a result of the Zeeman
coupling of the electron moment $\mu_0^{(2)}$ with the magnetic
field. For {\it ideal} Dirac cone, $\Phi_B$ is still equal to $\pm
\pi$ when the electron orbit surrounds the Dirac point.
\cite{berry} But if the electron band $\varepsilon_0({\bf k})$ is
not described by a strictly linear dependence in the whole energy
interval from the Dirac point to the Fermi energy $\varepsilon_F$,
this result for $\Phi_B$ generally fails, and the Berry phase may
differ from $\pi$. It is this possibility that was discussed in
Ref.~\onlinecite{taskin1}.

The functions ${\bf \Omega}_0({\bf k})$ and $\mu_0^{(2)}({\bf k})$
can be calculated \cite{zhetf,g-fac} using a ${\bf k}\cdot {\bf
p}$ Hamiltonian of electrons at a point ${\bf k}_0$ located near
the electron orbit. This calculation is also possible with an
effective two-band Hamiltonian that contains nonlinear terms in
${\bf k-k}_0$. \cite{g-fac1,prb11} Such the terms are usually
introduced into the Hamiltonian in order to take into account the
other bands different from the two considered explicitly. In the
magnetic field $H$ the effective two-band Hamiltonian should also
contain a term linear in $H$. This term is required for the
accurate calculation of $\mu_0^{(2)}({\bf k})$.

In Refs.~\onlinecite{Fu,ham} the effective two-band Hamiltonian
was found for the surface electrons of Bi$_2$X$_3$ family in the
vicinity of the center of their two-dimensional Brillouin zone,
\begin{equation} \label{8}
\hat H\!=bk^2\hat{\bf 1}\!+
v(k_y\hat\sigma_x-k_x\hat\sigma_y)\!+c(k_+^3+k_-^3)\hat\sigma_z\!+
\frac{\mu_B}{2}g_{sz}H_z\hat\sigma_z,
\end{equation}
where $b$, $v$, $c$, and $g_{sz}$ are some material-dependent
parameters; $\mu_B$ is the Bohr magneton, $k^2=k_x^2+k_y^2$;
$k_{\pm}=k_x \pm ik_y$; $\hat {\bf 1}$ is the unit matrix, and
$\hat \sigma_i$ are the Pauli matrices. At $H=0$ this Hamiltonian
describes the conduction and valence surface bands,
\begin{eqnarray} \label{9}
 \varepsilon_{c,v}({\bf k})&=&bk^2 \pm E, \\
 E^2 &\equiv& v^2k^2+4c^2k_x^2(k_x^2-3k_y^2)^2, \nonumber
\end{eqnarray}
which touch each other at the point $k=0$, and which  we denote by
the indexes $c$ and $v$. Note that these bands are not strictly
linear functions of $k$. They also contain the quadratic and cubic
corrections to the Dirac cone. The values of the parameters
defining Eq.~(\ref{8}) for the crystals of the Bi$_2$X$_3$ family
were estimated in Ref.~\onlinecite{ham}.

Using Hamiltonian (\ref{8}) and formulas (A2), (A3), (A7), (A8) of
our paper, \cite{g-fac} we calculate $\mu^{(2)}({\bf k})$ and
${\bf \Omega}({\bf k})$, e.g., for the conduction band $c$,
\begin{eqnarray}\label{11}
(\Omega_c)_{x,y}&=&\pm \frac{v^2k_{y,x}}{2E(E+2ck_x
 (k_x^2-3k_y^2))},\\
 \mu_{c}^{(2)}({\bf k})&=&\frac{ck_x(k_x^2-3k_y^2)}{E}\left(\frac{2v^2}{E\hbar }
-\frac{\hbar g_{zs}}{2m}\right),\ \ \ \ \label{12}
\end{eqnarray}
where $m$ is the electron mass, and the signs $+$ and $-$
correspond to $(\Omega_c)_{x}$ and $(\Omega_c)_{y}$, respectively.
Note that $\mu_{c}^{(2)}$ is proportional to the parameter $c$
defining the hexagonal warping term in Hamiltonian (\ref{8}).
Without this term, the electron moment lies in the $x$-$y$ plane,
\cite{Fu} and so it has no component along the magnetic field.

Using formulas (\ref{11}), (\ref{12}) and an expression for the
velocity $v_c=|\partial \varepsilon_c({\bf k})/\partial {\bf k}|$
that is directly obtainable from Eqs.~(\ref{9}), one can find
$\gamma$, calculating the integrals over the orbit $\Gamma$ in
Eq.~(\ref{7}). This orbit is defined by the condition
$\varepsilon_{c}({\bf k})=\varepsilon_F$ where the constant
$\varepsilon_F$ is the Fermi energy. It is significant that the
orbit is symmetric relative to the transformation $k_x\to -k_x$,
$k_y \to -k_y$, while the term $\mu_{c}^{(2)}$ is antisymmetric to
this transformation. Thus, the second integral in Eq.~(\ref{7}) is
always equal to zero.  Representing ${\bf \Omega}$ as a sum of the
symmetric ${\bf \Omega}^s\equiv [{\bf \Omega}({\bf k})+{\bf
\Omega}({\bf -k})]/2$ and antisymmetric ${\bf \Omega}^a\equiv
[{\bf \Omega}({\bf k})-{\bf \Omega}({\bf -k})]/2$ parts, we find
that only the antisymmetric part gives a nonzero contribution to
$\gamma$. A direct calculation with Eq.~(\ref{11}) yields
\begin{equation} \label{13}
 (\Omega_c^a)_{x,y}= \pm \frac{k_{y,x}}{2k^2}.
\end{equation}
This ${\bf \Omega}_c^a$ leads to $\Phi_B= -\pi$, and hence one
obtains $\gamma=0$ for any electron orbit in the conduction band
(the values $\gamma=1$ and $\gamma=0$ are equivalent in the
semiclassical limit). A similar calculation gives $\Phi_B= \pi$,
and $\gamma=0$ for the valence band. Interestingly, if the Dirac
cone in the Brillouin zone were shifted from the point ${\bf k}=0$
or were deformed asymmetrically, the Berry phase and the Zeeman
term in Eq.~(\ref{7}) would not have the universal values and
would depend on $\varepsilon_F$; see below.

In fact, the obtained result, $\gamma=0$, is exclusively caused by
the time-reversal symmetry of the Hamiltonian. This symmetry alone
dictates that any electron orbit is symmetric relative to the
transformation ${\bf k}\to -{\bf k}$, and that the Hamiltonian
should have the form
\begin{eqnarray} \label{14}
\hat H= h_0({\bf k})\hat {\bf 1}&+&\sum_{i=1}^3 h_i({\bf
k})\hat\sigma_i \nonumber \\
 &+&\frac{\mu_B}2 H_z [g_0({\bf
k})\hat {\bf 1}+ \sum_{i=1}^3 g_i({\bf k})\hat{\sigma}_i],
\end{eqnarray}
where arbitrary functions $h_0({\bf k})$, $h_i({\bf k})$ and
$g_0({\bf k})$, $g_i({\bf k})$ satisfy only the requirements:
$h_0({\bf k})=h_0(-{\bf k})$, $g_0(-{\bf k}) = - g_0({\bf k})$,
and $h_i({\bf k})=-h_i(-{\bf k})$, $g_i(-{\bf k})=g_i({\bf k})$.
With Eq.~(\ref{14}) we find the following expressions for ${\bf
\Omega}_c({\bf k})$ and $\mu_c^{(2)}({\bf k})$ generalizing
formulas (\ref{11}) - (\ref{13}):
\begin{eqnarray}\label{15}
{\bf \Omega}^a_c&=& \frac{1}{2(h_1^2+h_2^2)}\left(
h_2\frac{\partial h_1}{\partial {\bf k}}-h_1\frac{\partial
h_2}{\partial {\bf k}}\right) , \\
 {\bf \Omega}^s_c&=&-\frac{h_3}{\sqrt{h_1^2+h_2^2+h_3^2}}\,
{\bf \Omega}^a_c, \label{16} \\
  \mu_c^{(2)} &=&
-\frac{1}{2\hbar (h_1^2+h_2^2+h_3^2)}
\sum_{i,j,l=1}^3\varepsilon_{ijl}h_i\frac{\partial h_j}{\partial
k_x} \frac{\partial h_l}{\partial k_y} \nonumber \\
&-&\frac{\hbar}{4m}\left(g_0+\frac{1}{\sqrt{h_1^2+h_2^2+h_3^2}}
\sum_{i=1}^3g_ih_i\right), \label{16a}
\end{eqnarray}
where $\varepsilon_{ijl}$ is the completely antisymmetric unit
tensor with $\varepsilon_{123}=1$. A direct calculation shows that
rot${\bf \Omega}^a_c=0$ everywhere except the Dirac point for any
functions $h_1({\bf k})$, $h_2({\bf k})$. Hence, for the
calculation of the Berry phase, it is sufficient to consider
$\int_{\Gamma}{\bf \Omega}^a_cd{\bf k}$ in the immediate vicinity
of the Dirac point where these functions reduce to linear ones.
For linear functions $h_1({\bf k})$, $h_2({\bf k})$ one indeed
obtains $\Phi_B=\pm \pi$. As to the term $\mu_{c}^{(2)}({\bf k})$,
it is antisymmetric relative to the transformation ${\bf k}\to
-{\bf k}$ and does not contribute to $\gamma$. However, we
emphasize here that formulas (\ref{14}) - (\ref{16a}) are, in
principle, applicable to a general case (i.e., not only to the
Bi$_2$X$_3$ family of TI), and that these formulas can lead to
nontrivial values of $\gamma$ if the appropriate electron orbits
in some TI becomes asymmetric with respect to the point ${\bf
k}=0$.

The asymmetry of the orbit can appear when the external magnetic
field is not strictly perpendicular to the surface of TI, i.e.
when it has an in-plain component ${\bf H}_{\parallel}$. In this
case, e.g., Hamiltonian (\ref{8}) contains an additional term
\cite{Fu}
 \begin{equation}\label{20}
\delta \hat H=
g_{\parallel}\frac{\mu_B}{2}(H_x\sigma_x+H_y\sigma_y),
 \end{equation}
where $g_{\parallel}$ is the in-plain $g$ factor,
$g_{\parallel}\sim 1$. \cite{ham} This term shifts the Dirac cone
of the surface electrons in their Brillouin zone from the point
${\bf k}=0$ to the point ${\bf k}_*= g_{\parallel} (\mu_B/2v)[{\bf
H}\times {\bf i}_z]$,\cite{Fu} and the electron orbits become
asymmetric. This asymmetry generally leads to a nonzero value of
$\gamma$. However, in the first order in the small parameter
$g_{\parallel}\mu_B H_{\parallel} /\varepsilon_F$ the correction
to $\gamma$ vanishes for the case of Hamiltonian (\ref{8}) since
the angular dependence $\cos3\phi$ characteristic of the factor
$h_3({\bf k})$ in Eq.~(\ref{16}) differs from the angular
dependence $\cos\phi$ (or $\sin\phi$) that describes the
deformation of the electron orbits due to the field
$H_{\parallel}$. Here $\phi$ is the angle between ${\bf k}$ and
the $x$ axis. Due to this difference in the angular dependences,
the appropriate integrals over $\phi$ determining the correction
vanish. Thus, for the family Bi$_2$X$_3$ of TI the effect of the
in-plain magnetic field on $\gamma$ is negligible. This conclusion
is in an agreement with the experimental data
\cite{nat,ong,taskin} showing that the phase of the Shubnikov - de
Haas oscillations does not alter after applying the field
$H_{\parallel}$.

We now discuss the relation of the obtained results with the
published experimental data. In Ref.~\onlinecite{nat} it was shown
that in (Bi$_{1-x}$Sb$_x$)$_2$Se$_3$ the electron Landau levels
$\varepsilon_n$ at small $n$ can be well described by the formula
\begin{equation} \label{17}
\varepsilon_n=\pm\sqrt{\left(\frac{\mu_B}{2}g_{sz}H\right)^2+\frac{2ne
H v^2}{c\hbar}},
\end{equation}
where $g_{sz}\approx 50$. This formula gives the exact Landau
levels of Hamiltonian (\ref{8}) at $b=c=0$. Let us rewrite
expression (\ref{17}) in the form of quantization condition
(\ref{1}),
 \begin{equation}\label{18}
S(\varepsilon_n)=\frac{2\pi e H}{\hbar c}\left(n+
\frac{Hg_{sz}^2\mu_B^2c\hbar}{8ev^2} \right),
 \end{equation}
where $S(\varepsilon)=\pi k^2=\pi (\varepsilon/v)^2$ is the area
of the electron orbit in the ${\bf k}$-space, and consider the
case of large $n$ (i.e., of relatively small $H$). This case just
corresponds to the semiclassical approximation used in our
calculations. Now the second term in the right hand side of
formula (\ref{18}) is small as compared to the first one, and we
may express $H$ via $n$ as follows: $H\approx \hbar c S/(2\pi e
n)$. Inserting this expression for $H$ in the second term of
Eq.~(\ref{18}), we arrive at
 \begin{equation}\label{19}
S(\varepsilon_n)\approx \frac{2\pi e H}{\hbar c}\left(n+ \frac{S
g_{sz}^2\mu_B^2c^2\hbar^2}{16\pi e^2v^2 }\frac{1}{n} \right).
 \end{equation}

Strictly speaking, the right hand side of the quantization
condition (\ref{1}) is a series in $1/n$ [i.e., it has the form
$n+{\rm const}+(1/n)+ ...$] in which only the first two terms,
$n+\gamma$, are usually kept. It is this cut of the series that
yields Eq.~(\ref{1}). Formula (\ref{19}) shows that spectrum
(\ref{17}) in the semiclassical limit ($n\gg 1$) leads to
$\gamma=0$ in an agreement with our results obtained above and
that in contrast with the case of the usual electron gas, a finite
$g_{sz}$ in Eq.~(\ref{17}) gives a contribution only to the term
of the order of $1/n$. In other words, the nonzero values of
$\gamma$ found in the experiments, \cite{nat,ong,taskin} are due
to the difference of the exact spectrum from the semiclassial one
at not too large values of $n$. This conclusion is supported by an
analysis of the Shubnikov - de Haas oscillations carried out by
Taskin and Ando \cite{taskin1} for various TI. Note that in
graphene one has a relatively small value of $g_{sz}\sim 2$, and
the semiclassical spectrum defined by Eq.~(\ref{1}) practically
coincides with the exact spectrum (\ref{17}) even at $n=0$ and
$1$. The large value of $g_{sz}\approx 50$ for
(Bi$_{1-x}$Sb$_x$)$_2$Se$_3$ is probably due to a relatively small
gap between the Dirac-point energy and the energies of some other
surface bands at ${\bf k}=0$ for this TI.

In summary, within the semiclassical approach, we derive formula
(\ref{7}) for the constant $\gamma$ that enters the quantization
condition (\ref{1}) and defines the phase $\varphi=-2\pi\gamma$ of
the Shubnikov - de Haas oscillations in TI. This $\gamma$ is
determined by the Berry phase of an electron orbit and by a part
of the electron moment averaged over the orbit. Since the Dirac
point lies at ${\bf k}=0$ for Bi$_2$X$_3$ family of TI, we find
that the Berry phase is equal to $\pi$, while the averaged moment
is zero, and hence one always has $\gamma=0$ for this family of
TI. The nonzero values of $\gamma$ found in some experiments
appear to be due to the difference of the exact electron spectrum
from the semiclassial one at not too large quantizing numbers $n$.

{}

\end{document}